\documentclass[letterpaper,twocolumn,10pt]{ilass}
\def\thepage{}

\usepackage{amssymb}
\usepackage{amsmath}
\usepackage{nomencl}
\usepackage[round]{natbib}

\usepackage{ifpdf}

\ifpdf
\usepackage[pdftex]{graphicx}
\else
\usepackage{graphicx}
\fi

\title{\vspace{-0.25in} 
        \large {\bf High-Fidelity Multi-Scale Simulation of Swirled Air-blast Atomization \\ 
with Comparison against Experiments
} }
\author{\large Lorenzo Bruni\textsuperscript{1,2} and Olivier Desjardins\textsuperscript{1}\footnote{Corresponding Author: olivier.desjardins@cornell.edu} \\
        \textsuperscript{1} \large Sibley School of Mechanical and Aerospace Engineering \\ Cornell University \\ Ithaca, NY 14853 USA
        \large  \\ 
        \textsuperscript{2} \large Department of Civil and Industrial Engineering \\ Università di Pisa \\ Pisa, 56126 Italy
        \large }

\date{ \normalsize  \centerline{\bf Abstract} \vspace{0.05in}
 \begin{minipage}{6.5in}
 \normalsize In liquid-fueled combustion systems, optimization of the fuel atomization process is critical to reducing fuel consumption and lowering pollutant emissions. Accurate and efficient computational modeling of liquid atomization can open the door to spray optimization, however it presents a significant challenge to modelers due to the extremely complex flow field and wide range of length and time scales involved. In this work, a multi-scale and multi-domain simulation strategy is used to model end-to-end the turbulent spray produced by a swirled two-fluid coaxial atomizer, a device that utilize a high-speed swirled gas stream to destabilize a co-flowing low-speed liquid, widely used in systems such as fuel injectors. Our computational method relies on sub-grid scale modeling; in particular, we will introduce a thin structure break-up model to account for topology changes, converting thin liquid structures into spherical Lagrangian particles. With such simulations, the impact of swirl on the break-up process can be analyzed by varying the swirl ratio, and we aim to quantitatively validate our simulations against experiments at identical operating conditions, including drop size statistics.
 \end{minipage} \vspace{-0.25in}
}

\begin{document}

\ifpdf
\DeclareGraphicsExtensions{.pdf, .jpg}
\else
\DeclareGraphicsExtensions{.eps, .jpg}
\fi

\maketitle

\clearpage 
\pagenumbering{arabic}
\setcounter{page}{2}

\section*{Introduction} 
Liquid atomization is employed in a wide range of engineering applications such as pharmaceutical processes and agricultural systems. In addition, it is also used for combustion systems, and as such optimizing fuel atomization is of crucial importance in the context of the current climate crisis. Improved computational models of atomization can help develop strategies for reducing fuel consumption as well as pollutant emission. This work focuses on high-fidelity multi-scale modeling of air-blast atomization, in which a two-fluid coaxial atomizer is considered. This atomizer brings in contact a low-speed liquid jet with a high-speed gas phase that peels off structures from the liquid surface, generating a distribution of spray droplets in the process. In particular, this work focuses on the effect of a swirling gas phase to further enhance liquid destabilization. Validation against various experimental data is planned, such as visible light back-lit imaging and synchrotron X-ray radiography. The simulation presented herein is performed using the open-source code NGA2.

\section*{Atomizer Design}
The atomizer, shown in Figure 1, is a canonical airblast nozzle that has been studied extensively \citep{Machicoane2019}. It is composed of a liquid needle from which liquid is issued, surrounded by a concentric annular gas nozzle connected to a converging plenum shaped from cubic splines. The gas enters from four axial ports of diameter equal to 16 mm that are perpendicular to the axis of the injector. Four additional ports of diameter 7.1 mm are offset by 30 mm with respect to the main axial ports, allowing to introduce swirl in the gas phase. Further details about the atomizer exit dimensions are provided in Figure 2 and Table 1. The gas enters the plenum and accelerates as it flows towards the nozzle exit where, at a high speed, it comes in contact with the low-speed liquid phase, contributing to its break-up in the atomizing region. At the conditions considered in this work, the liquid comes out of the liquid needle with a fully developed Poiseuille profile.
\begin{figure}[h]
\begin{center}
\includegraphics[width=0.49\textwidth]{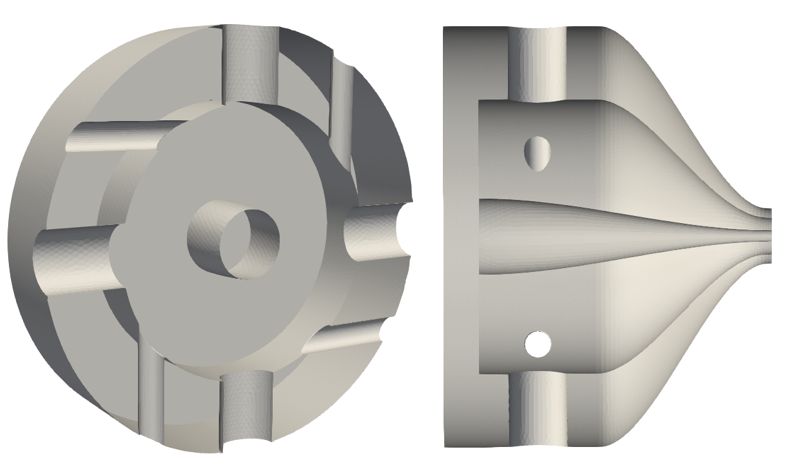}
\end{center}
\caption{Atomizer configuration illustrated through two cut views.}
\label{Figure 1} 
\end{figure}

\begin{figure}[h!]
\begin{center}
\includegraphics[width=3in]{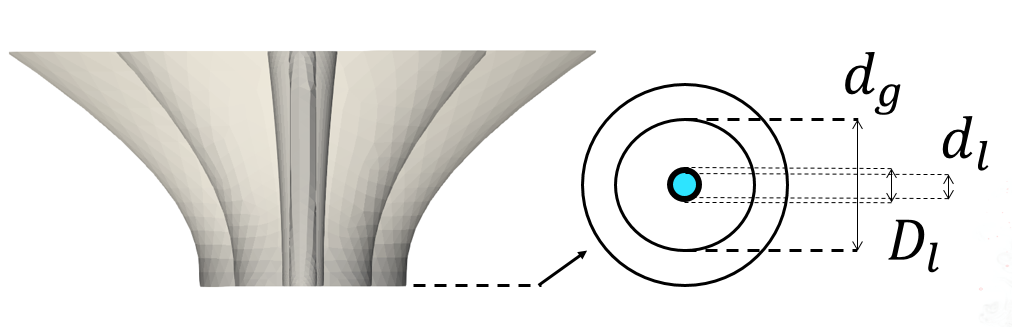}
\end{center}
\caption{Zoom on the atomizer exit.}
\label{Figure 2} 
\end{figure}

\begin{table}[h!]
\begin{center}
\begin{tabular}{| r | r |} \hline
$d_l$ & 2 mm  \\ \hline
$D_l$ & 3 mm  \\ \hline
$d_g$ & 10 mm \\ \hline
\end{tabular}
\end{center}
\caption{Atomizer exit dimensions.}
\label{Table 1}
\end{table} 

Table 2 shows the most relevant non-dimensional parameters for the simulation: gas Reynolds number ($Re_g=4Q_T/\sqrt{4\pi A_g}\nu_g$), liquid Reynolds number ($Re_l=\rho_l U_l d_l/\mu_l$), gas-to-liquid dynamic pressure ratio ($M=(\rho_g U_g^2)/(\rho_l U_l^2)$), Weber number based on the liquid jet diameter but gas density and slip velocity between the two phases ($We=\rho_g(U_g-U_l)^2 d_l/\sigma$), density ratio ($\rho^\star$) and viscosity ratio ($\mu^\star$),  where $A_g$ is the gas flow-through exit area, $\rho_s$ is the fluid density, $\mu_s$ is the dynamic viscosity, $\nu_s$ is the kinematic viscosity, $\sigma$ is the surface tension coefficient, $U_s$ is the bulk velocity and subscript $s$ indicates liquid ($s=l$) and gas ($s=g$) quantities, respectively. $Q_{sw}$ is the swirling gas flow rate through the four offset gas ports and $Q_{ax}$ is the axial gas flow rate through the four axial gas ports. $Q_T=Q_{ax}+Q_{sw}$ and $Q_l$ are the total and the liquid flow rates, and the corresponding swirl ratio is $SR=Q_{sw}/Q_{ax}=75\%$. Table 3 summarizes a few relevant dimensional properties for the simulation.
\begin{table}[h]
\begin{center}
\begin{tabular}{|r|r|r|r|r|r|} \hline
$Re_g$ & $Re_l$ & $M$ & $We$ & $\rho^\star$ & $\mu^\star$ \\ \hline
21400 & 1200 & 6.4 & 39.1 & 815 & 65  \\ \hline
\end{tabular}
\end{center}
\caption{Simulation non-dimensional parameters.}
\label{Table 2}

\begin{center}
\begin{tabular}{|r|r|r|r|r|} \hline
$\rho_l$ (kg/m$^3$) & $Q_{ax}$ & $Q_{sw}$ & $Q_T$ & $Q_l$   \\
\hline
1000 & 85.7 & 64.3 & 150 & 0.099 \\ 
\hline

\end{tabular}
\end{center}
\caption{Relevant properties of the simulation. Volume flow rates are in L/min.}
\label{Table 3}
\end{table}

\section*{Geometry Setup}
The geometry of the atomizer based on CAD files is  converted to a surface mesh representation and read into the NGA2 code. The Cartesian flow solver mesh is intersected with the surface mesh, and a second-order accurate sharp solid volume fraction field is built. This information is then used in a volumetric direct forcing immersed boundary scheme loosely based on the approach of \cite{Bigot2014}.

\section*{Simulation Setup}
The characteristics of the atomization process are expected to depend on the details of the gas flow that is issued from the nozzle, and as such the flow in the nozzle needs to be predicted accurately. To account for the widely different length scales that characterize the nozzle plenum compared to the smallest drops in the atomizing region, we make use of two computational domains that are coupled to one another via their boundary conditions. 

\subsection*{\underline{Domain 1 -- Nozzle}}
A first computational domain focuses on the internal gas flow in the nozzle. That domain spans the entire nozzle, from the inlet pipes to beyond the nozzle exit, as indicated by the red outline in Figure 3. In this domain, the liquid flow in the central pipe is ignored, so that we only need to solve for the gas flow, allowing to use a fast single-phase large-eddy simulation (LES) solver. Gas ports are configured as Dirichlet bulk inflow conditions, while the nozzle exit is set up as a Neumann outflow condition. A $600^3$ uniform computational mesh is considered, leading to a resolution of $\Delta x= 140~\mu$m.

\subsection*{\underline{Domain 2 -- Atomization}}
A second computational domain focuses on the modeling of interface deformation and topology change of the liquid as it undergoes atomization. In this domain, NGA2's high-resolution VOF-based two-phase LES is employed. Inside each multiphase computational cell, the interface is represented as a plane using the piece-wise linear interface construction (PLIC), with the plane normal calculated using ELVIRA \citep{Pilliod2004}. The curvature of the interface is calculated using parabolic surface fits. The domain starts one centimeter before the nozzle exit, and extends 3 centimeters downstream. The domain width is 4 centimeters in both lateral directions, as illustrated with the pink outline in Figure 3. The liquid inlet is based on a Poiseuille solution provided through a Dirichlet condition, and the gas inlet is provided as a time-varying Dirichlet condition obtained by interpolating the solution from the nozzle domain in a time-accurate fashion using NGA2's parallel coupler. The outflow is again a Neumann condition, while the domain sides are set up as slip conditions that allow for gas entrainment to enforce global mass conservation. The mesh resolution is uniform with $\Delta x=100~\mu$m.
 
\begin{figure}[h]
\begin{center}
\includegraphics[width=3in]{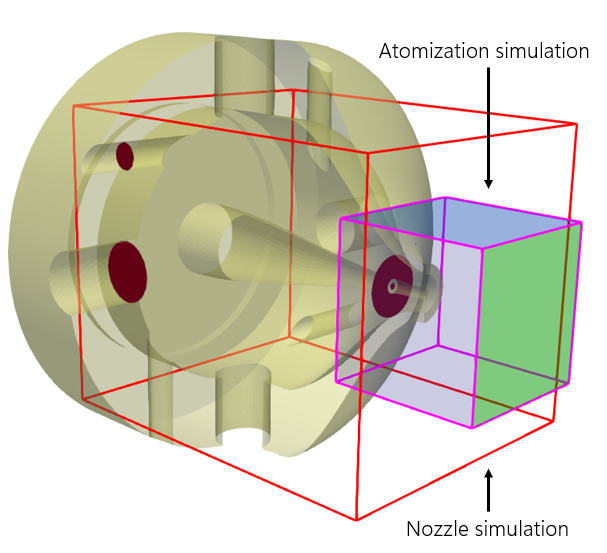}
\end{center}
\caption{Simulation setup with the two coupled computational domains. Inflow pipes for axial and swirled flow and the coupling plane are highlighted in red. The green face in the atomization simulation in the outflow condition and the blue side faces are slip conditions that allow for entrainment.}
\label{Figure 3} 
\end{figure}

\begin{figure*}[!ht]
\begin{center}
\includegraphics[width=0.95\textwidth]{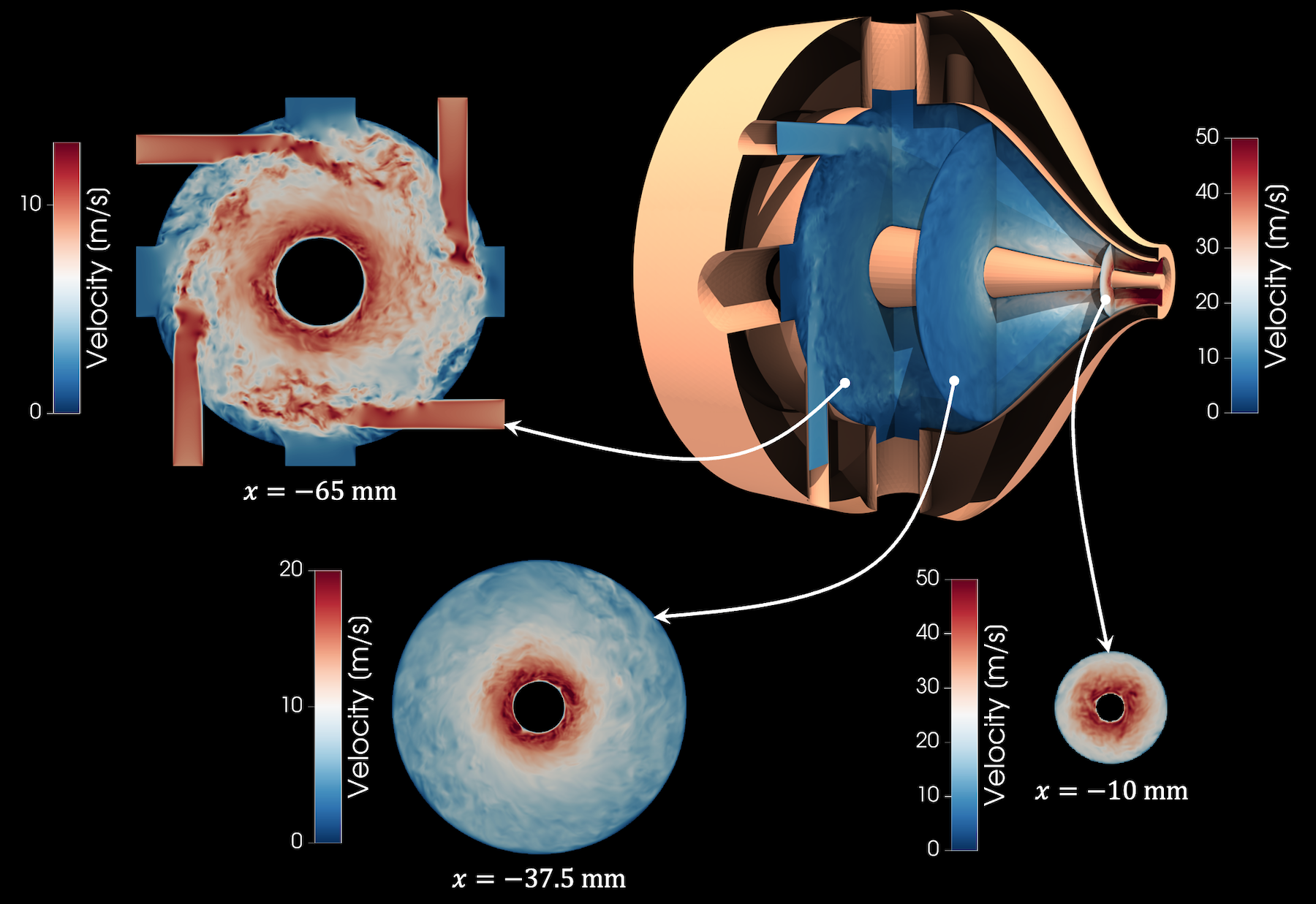}
\end{center}
\caption{Instantaneous velocity magnitude throughout the atomizer on various cut planes.}
\label{Figure 4}
\end{figure*}

\section*{Simulation Results}
The simulation is run from fluid at rest for about 200 ms. Examination of the nozzle flow indicates that it takes about 100 ms for swirl to start developing throughout the nozzle, and about 150 ms for the nozzle flow to reach statistically stationary conditions. Figure 4 illustrates the instantaneous turbulent flow field at various locations throughout the nozzle. The low speed of the nozzle inlets leads to a laminar behavior over all eight gas ports. However, the interaction between the inlet jets inside the plenum leads to a rapid transition to turbulence, as visible in the first cut plane at $x=-65$ mm on the left of Figure 4. As the nozzle contracts, the gas flow rapidly accelerates, reducing turbulence intensity in the process. The flow reaches its maximum speed close to the liquid needle, as a strong swirling vortex core develops inside the atomizer.

\begin{figure*}[!ht]
\begin{center}
\includegraphics[width=0.75\textwidth]{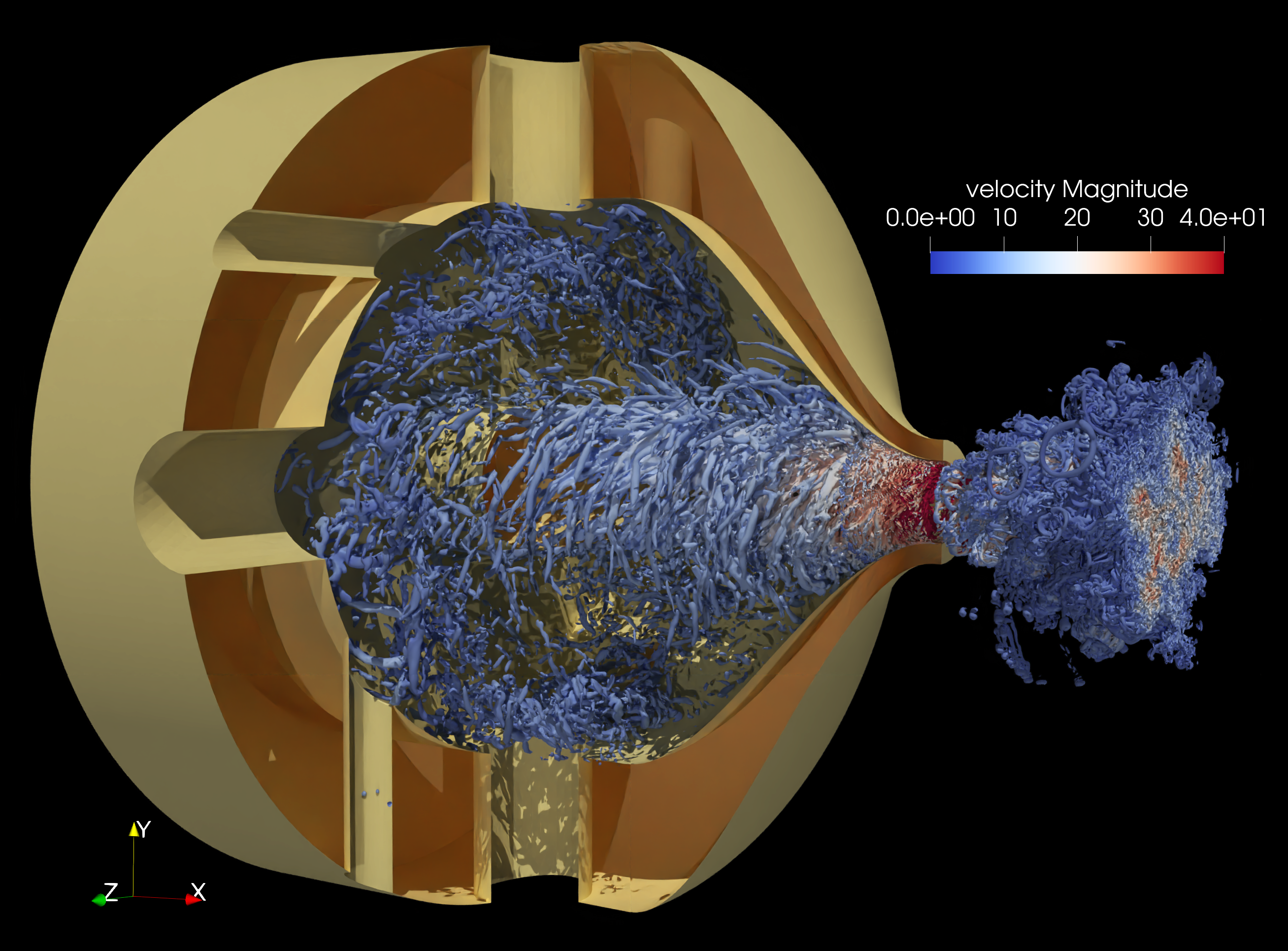}
\includegraphics[width=0.75\textwidth]{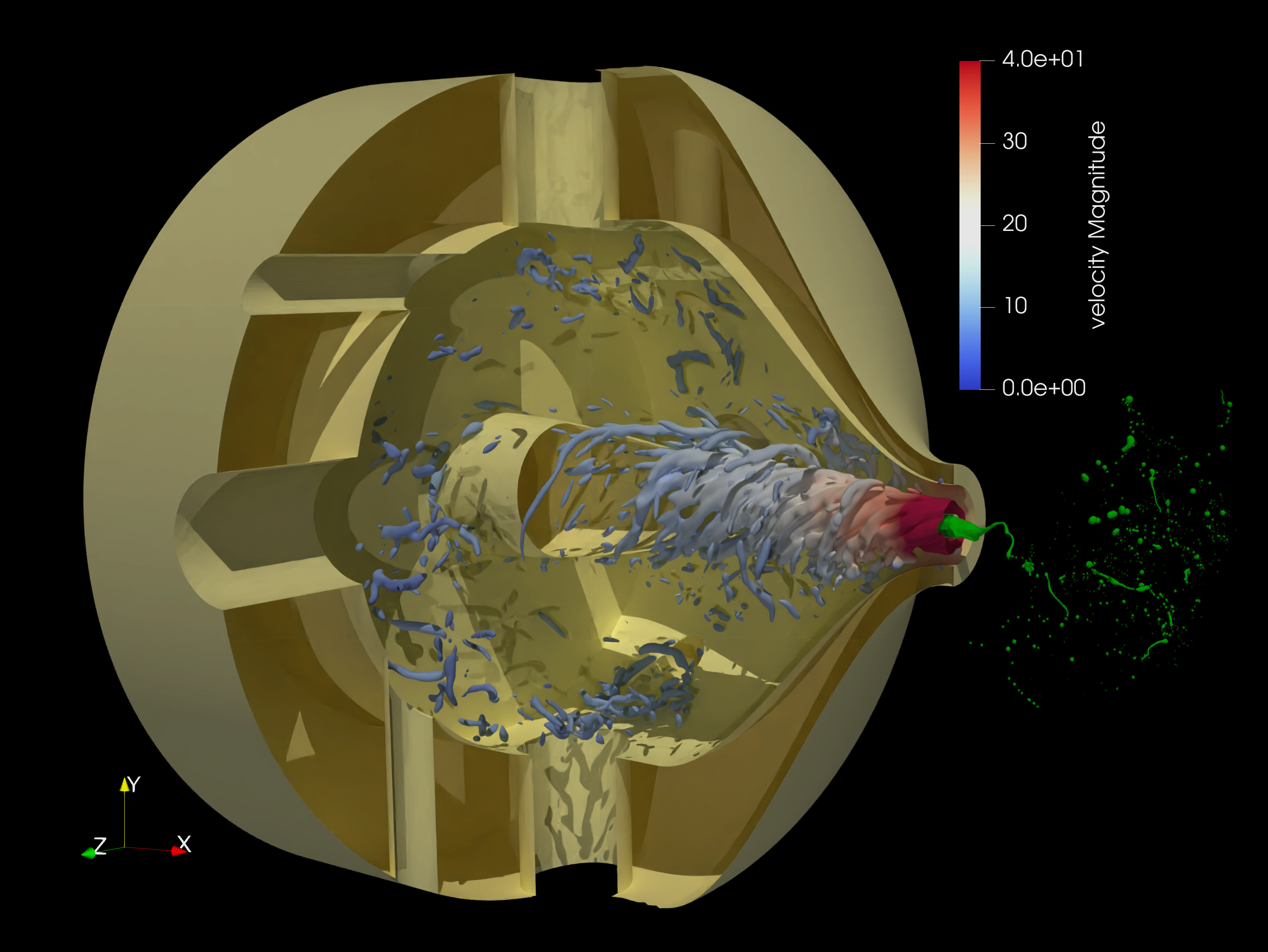}
\end{center}
\caption{Visualization of an isosurface of the Q criterion (top) and of the low-pass-filtered Q criterion (bottom), colored with the velocity magnitude. In the atomization region, the bottom image shows the liquid-gas interface in green instead of the isosurface of the Q criterion.}
\label{Figure 5} 
\end{figure*}

To better visualize the vortical motion in the nozzle, Figure 5 presents a three-dimensional visualization of an isosurface of the Q criterion, the second invariant of the velocity gradient tensor. The top view in Figure 5 highlights the extreme complexity of the turbulent flow field, both inside and outside the nozzle. Since Q is based on the velocity gradient tensor, it inherently scales with the smallest scales of the flow, thereby making it challenging to visualize large-scale and energy-containing vortical motion in high Reynolds number flows. To alleviate this issue, the bottom view in Figure 5 presents an isosurface of a low-pass-filtered Q criterion, which highlights more clearly the formation of a strong vortical core along the liquid needle. This vortex core directly interacts with the liquid jet (shown in green) in the atomization region, forcing it to undergo intense swirling.

\begin{figure}[!ht]
\begin{center}
\includegraphics[width=3in]{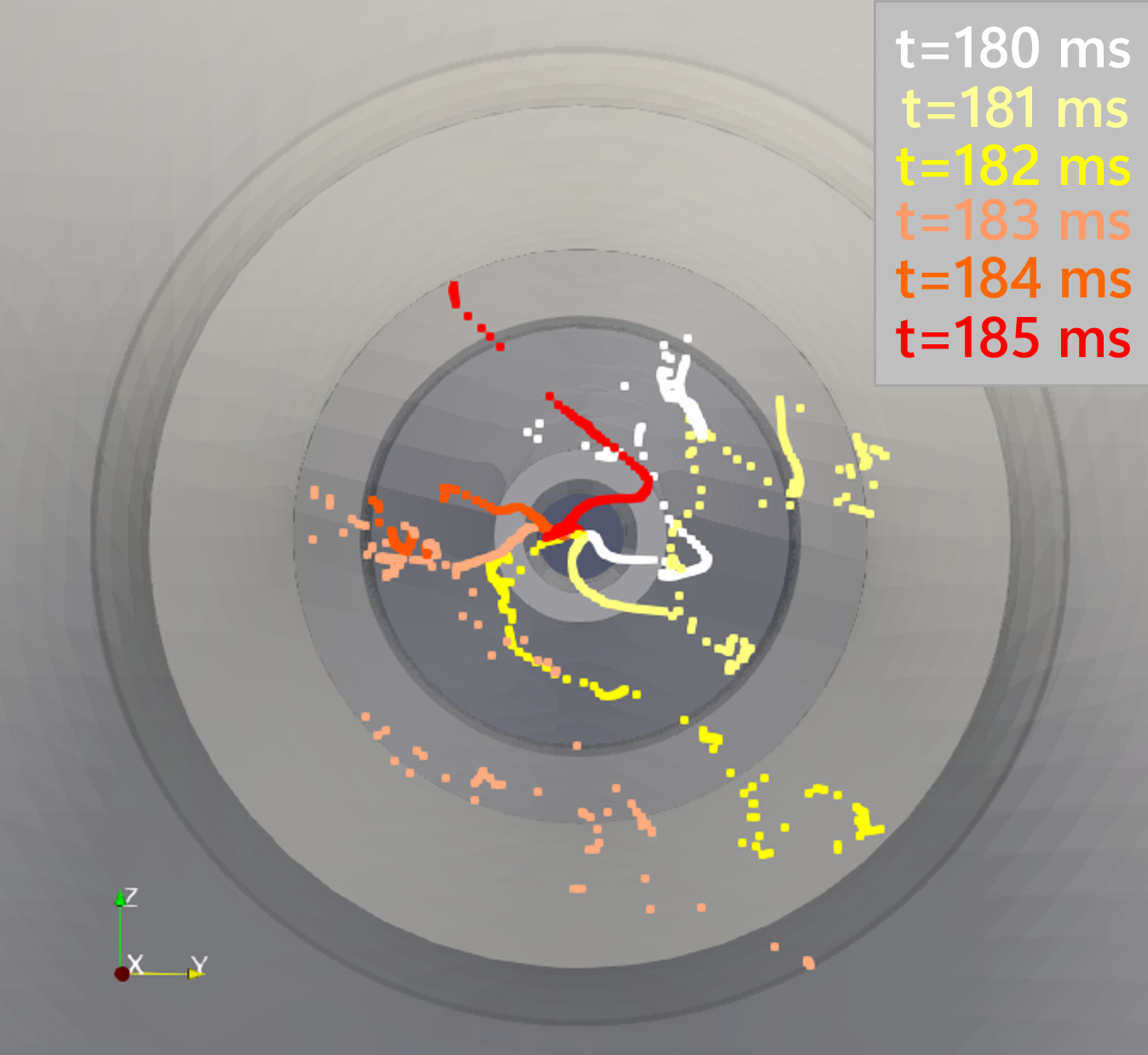}
\end{center}
\caption{Visualization of the liquid core swirling dynamics.}
\label{Figure 6}
\end{figure}

To visualize better the swirling liquid dynamics, we extract a skeleton of the liquid core in the atomization region. Using a connected component labeling technique, the liquid core is extracted and the $y-z$ location of its barycenter is analyzed as a function of the downstream distance $x$. Figure 6 presents the extracted liquid core skeleton at 6 successive times chosen 1 ms apart from one another, clearly highlighting the swirling behavior of the liquid jet. 

\section*{Qualitative Comparisons against Experiments}
While quantitative comparisons against experiments are planned, this paper presents a preliminary discussion of qualitative comparisons against experimental back-lit imaging data presented by \cite{Kaczmarek2022}. Figure 7 presents a side-by-side view of the simulated and experimental liquid jet on the same scale, at the same flow conditions, with similar time between the images of 1 ms. In both experimental and simulation images, we can see the process of the formation and break-up of a liquid bag. This structure formation is one of the most frequently recurring processes of the jet, and they seem to be in excellent qualitative agreement between simulations and experiments. 
\begin{figure}[!ht]
\begin{center}
\includegraphics[width=3in]{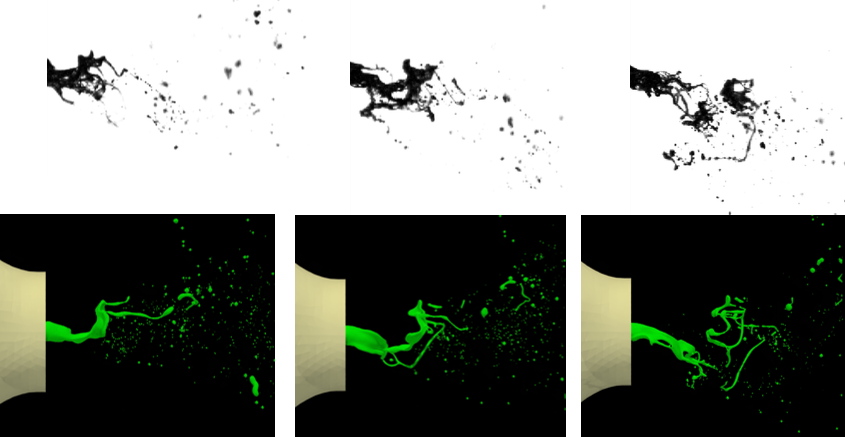}
\end{center}
\caption{Comparison of a bag formation and break-up between experiment (top) and simulation (bottom), for the same conditions, at the same scale. Images are shown with 1 ms separation.}
\label{Figure 7}
\hspace{1in}
\end{figure}

\section*{Assessment of Instabilities}

As the liquid-gas interface emerges from the liquid needle, it becomes first susceptible to a Kelvin--Helmholtz-type instability due to high level of shear imposed by the gas. Such an instability is visible in both experiments and simulations, as illustrated in Figure 8. Note that the same scale is used for both simulation and experimental images, highlighting the agreement between the observed wavelengths. To compare against theoretical expectations, we can make use of the most unstable expected wavelength given by
\begin{equation}
\lambda_{KH}=\frac{2\pi}{1.5}\delta \left(\frac{\rho_l}{\rho_g}\right)^\frac{1}{2}
\end{equation}
as employed for example by \cite{Marmottant2004}, where $\delta$ is the vorticity thickness.
Analyzing the gaseous flow at the nozzle exit, the vorticity thickness can be estimated to be about $0.09$ mm, leading to a theoretical prediction of $\lambda_{KH} \approx 10$ mm,  which is about four times larger than the value observed in simulations and experiments (which exhibit $\lambda_{KH} \approx 2.3$ mm). Note that a similar level of disagreement was reported by \cite{Marmottant2004} between their experiments and theoretical predictions.

\begin{figure}[!ht]
\begin{center}
\includegraphics[width=3in]{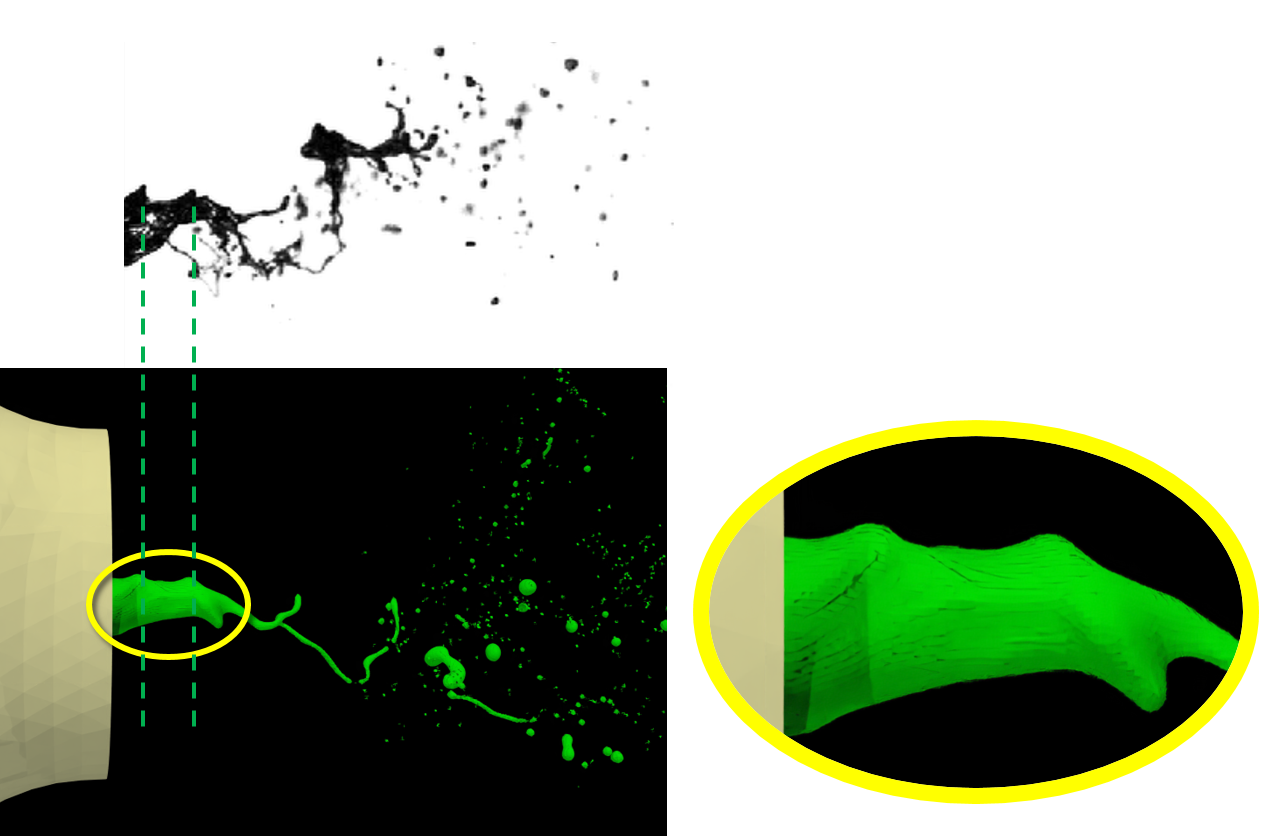}
\end{center}
\caption{Kelvin-Helmholtz-type instabilities in experiments and simulations.}
\label{Figure 8}
\end{figure}

\begin{figure}[!ht]
\begin{center}
\includegraphics[width=2in]{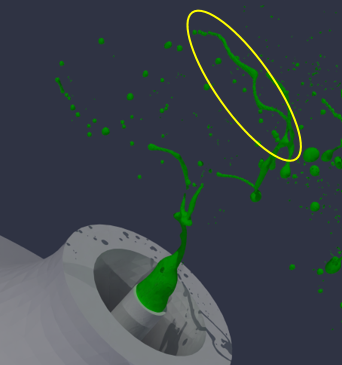}
\end{center}
\caption{Instability developing along the length of a ligament leftover from the break-up of a liquid bag.}
\label{Figure 9}
\end{figure}

As the liquid jet breaks up further, long ligaments are formed that exhibit very noticeable instabilities leading to undulations along the ligament length, as illustrated in Figure 9. Some of these ligaments appear to be exposed to a strong gaseous cross-flow: as a result, the high density liquid is being accelerated due to drag into a low density gas, which suggests that a Rayleigh--Taylor instability might causing the wavy pattern on the ligament. To test this hypothesis, we evaluate the most unstable wavelength due to Rayleigh--Taylor under the assumption that aerodynamic drag on a liquid cylinder is the cause of acceleration, leading to 
\begin{equation}
\lambda_{RT}=2\pi\sqrt{\frac{3\sigma}{|\rho_l-\rho_g|A_\text{drag}}}
\end{equation}
where
\begin{equation}
A_\text{drag}=\frac{F_D}{m_{lig}}=\frac{2C_D\rho_g U_\text{slip}^2}{\pi\rho_l d_{lig}}.
\end{equation}
In this expression, $F_D$ is the drag force, $m_{lig}$ is the mass of the ligament, $C_D$ is the drag coefficient, $U_\text{slip}$ is the relative velocity between gas and ligament, and $d_{lig}$ is the ligament diameter. To estimate these values, we look for the ligament in Figure 9 in a snapshot of the flow 1 ms earlier, i.e., prior to the development of the instability. Figure 10 shows a view of the gaseous cross-flow impacting that ligament, confirming that the approximation of a flow on a cylinder is plausible. From this image, we estimate $U_\text{slip} \approx 28$ m/s and $d_{lig} \approx 0.1$ mm. These values correspond to a gas Reynolds and Weber number for the ligament sub-system of 190 and 1.33, respectively. At that Reynolds number, we can estimate $C_D$ for a cylinder to be $C_D \approx \pi/2$, leading to an acceleration due to drag of $5000$ m/s$^2$ and $\lambda_{RT} \approx 1$ mm, which matches well with the wavelength that can be extracted from our simulation.

\begin{figure}[!ht]
\begin{center}
\includegraphics[width=2.5in]{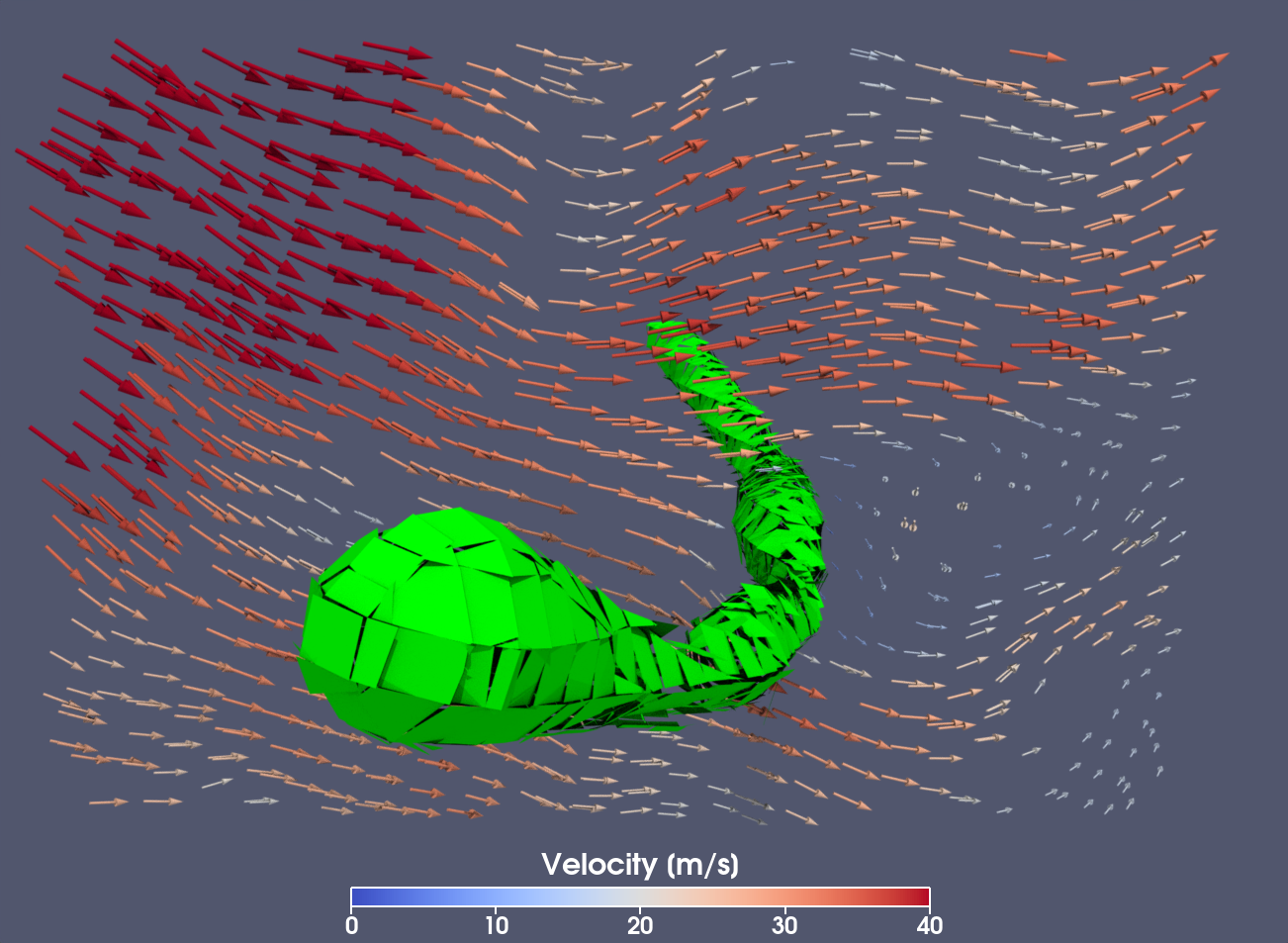}
\end{center}
\caption{Flow field around the ligament in Figure 9 before the development of the instability.}
\label{Figure 10}
\end{figure}

Nevertheless, we note that other mechanisms could be causing ligament waviness. First, the Rayleigh--Plateau instability is expected to be active on these ligaments and would lead to a most amplified wavelength of about $\lambda_{RP} \approx 9 R_{lig}$, which is of the same order of $\lambda_{RT}$. Moreover, other sources of acceleration are present on the ligament and could trigger Rayleigh--Taylor instabilities: in particular, in the context of our swirling flow, the spinning ligament experiences a centrifugal acceleration on the order of $1000$ m/s$^2$ as the liquid core revolves around the axis about every 6 ms. Again, this leads to a similar value of the most unstable wavelength. Figure 11 highlights an instance in the simulation were two ligaments develop waviness as they spin around the central axis.

\begin{figure}[!ht]
\begin{center}
\includegraphics[width=3in]{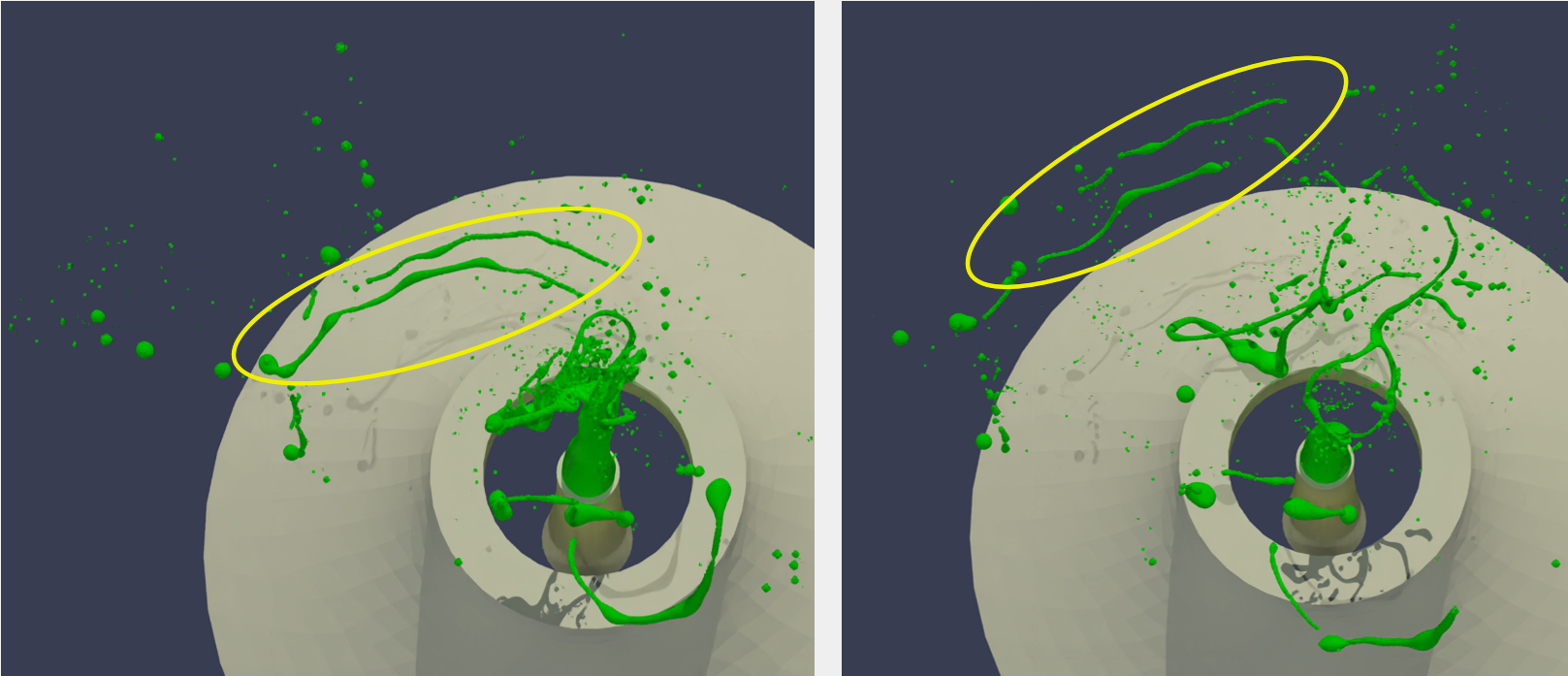}
\end{center}
\caption{Unstable ligaments undergoing circular motion.}
\label{Figure 11}
\end{figure}

\section*{Drop Size Statistics}
Finally, we analyze the drop size statistics by extracting the drop diameter probability distribution function (pdf) for all droplets passing through the exit plane of the atomization domain. To that end, a connected component labeling algorithm is employed to identify separated liquid structures, then the volume of these liquid structures is measured and an equivalent diameter is extracted. While no specific attention is given to the sphericity of the liquid structures, we explicitly exclude droplets with a diameter below the mesh resolution, as we expect them to be a result of numerical errors more so than physical break-up. Note that the mass of the excluded droplets represents about 0.01\% of the total mass of the spray. Figure 12 shows the resulting droplet diameter pdf. We observe that the probability decreases rapidly for droplets smaller than $2\Delta x$, a behavior that can be attributed to the numerical surface tension inherent to the use of ELVIRA as our reconstruction scheme.

\begin{figure}[!ht]
\begin{center}
\includegraphics[width=3.1in]{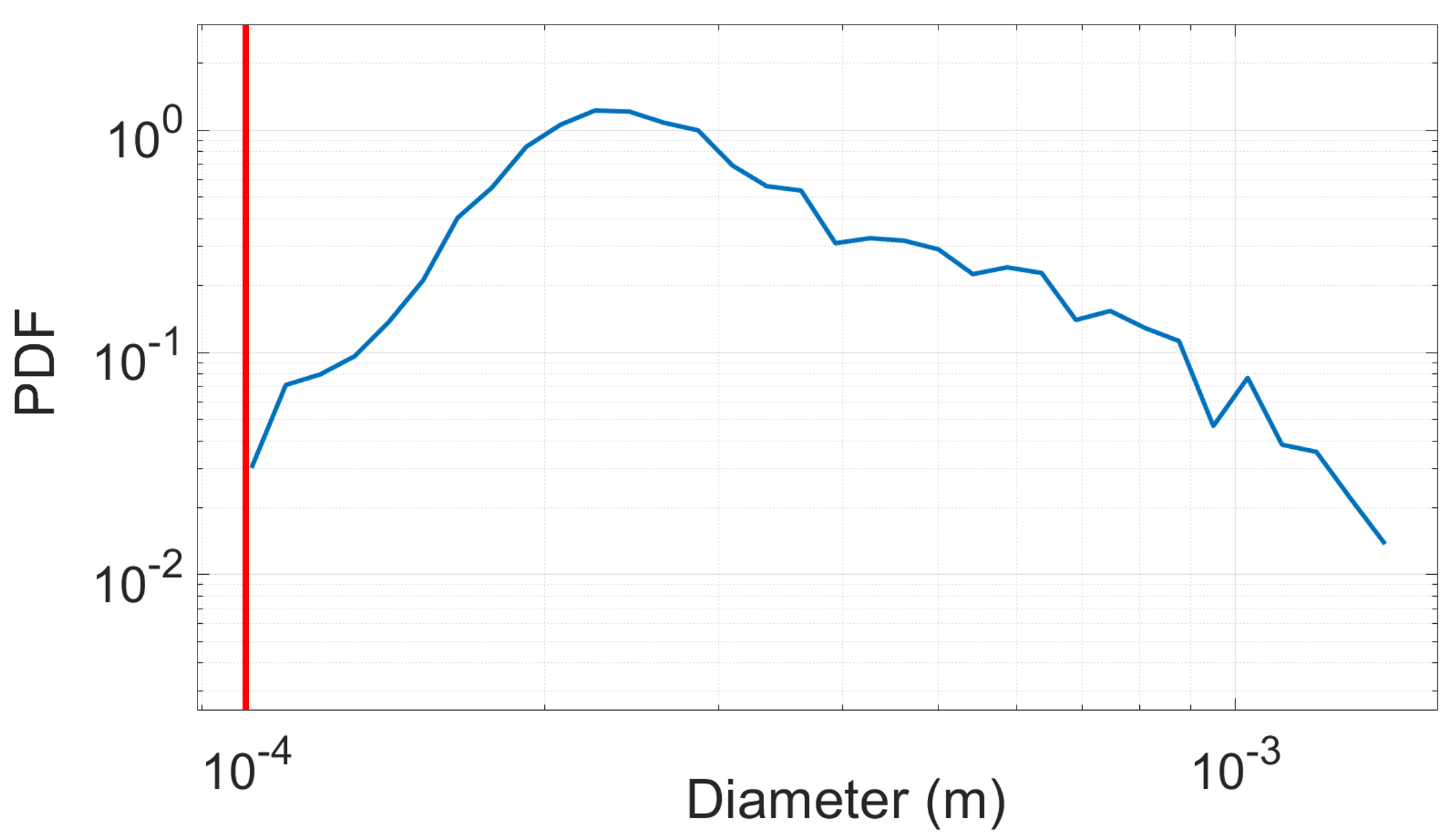}
\end{center}
\caption{Drop size pdf in blue. The red vertical line indicates the mesh size.}
\label{Figure 12}
\end{figure}

\section*{Conclusion}
In this study, we present a multi-domain simulation of a swirled airblast atomizer that generates computational predictions in good qualitative agreement with experiments. Our approach relies on two coupled domains, each of them considering a different region and scale of the flow. The first domain is used to represent the flow inside the atomizer, whereas the second considers the atomization region. We extract the liquid core to show its spinning behavior, and analyze the potential instabilities mechanisms in the flow. Finally, the drop size statistics are extracted and found to be strongly impacted by the mesh size, as could be expected from the computational techniques used in the model. Future work will involve employing a subgrid scale break-up model based on the tracking of thin liquid structures, and more extensive quantitative comparisons against experiments.

\section*{Nomenclature}
\noindent $d$ \hspace{0.3in} {Inner diameter} \\
$D$ \hspace{0.23in} {Outer diameter} \\
$U$ \hspace{0.25in} {Bulk velocity} \\
$Q$ \hspace{0.26in} {Flow rate} \\
$\nu$ \hspace{0.29in} {Kinematic viscosity} \\
$\mu$ \hspace{0.29in} {Dynamic viscosity} \\
$\rho$ \hspace{0.3in} {Density} \\
$\sigma$ \hspace{0.29in} {Surface tension} \\
$Re$ \hspace{0.20in} {Reynolds number} \\
$We$ \hspace{0.15in} {Weber number} \\
$\lambda$ \hspace{0.28in} {Wavelength} \\
$C\textsubscript{D}$ \hspace{0.18in} {Drag coefficient} \\
$\delta$ \hspace{0.31in} {Vorticity thickness} \\

\subsection*{Subscripts}
\noindent $g$ \hspace{0.36in}Gas \\
$l$ \hspace{0.4in}Liquid
\vspace{0.5in}

\bibliographystyle{ilass}

\end{document}